# MahaTTS: A Unified Framework for Multilingual Text-to-Speech Synthesis

Jaskaran Singh, Amartya Roy Chowdhury, Raghav Prabhakar, Varshul C. W.
{jaskaran, amartya.roy, raghav.prabhakar, varshul}@dubverse.ai

**Abstract**

*Current Text-to-Speech models pose a multilingual challenge, where most of the models traditionally focus on English and European languages, thereby hurting the potential to provide access to information for many more people. To address this gap, we introduce **MahaTTS-v2** a multilingual multispeaker text-to-speech (TTS) system that has excellent multilingual expressive capabilities in indic languages. The model has been trained on around $\sim 20K$ hours of data specifically focused on Indian languages. Our approach leverages wav2vec2.0 tokens for semantic extraction, and a Language Model (LM) for text-to-semantic modeling. Additionally, we have used a Conditional Flow Model (CFM) for semantics to Mel spectogram generation. The experimental results indicate the effectiveness of the proposed approach over other frameworks. Our code is available at https://github.com/dubverse-ai/MahaTTSv2*

## 1. Introduction

In recent years research has shown great advancements in Text-to-Speech (TTS), but most of these systems are in English and or chinese, With a few in multilingual settings, trained on large speech data delivering exceptional intonation and quality speech. When trained on large data, there are different emergent properties cross linguality and multilingualism without any hard specific data collected for. Tortoise TTS [1] used a VQVAE [14] model to quantize the speech into semantic tokens/units, which are used to train a text to semantic token model based on GPT2 [11] architecture with different learned positional encoding and classification heads for text and semantic tokens, a second model is used to go from semantic tokens to melspectogram which is then converted into audio waveform using a univnet [5] vocoder. Tortoise-tts work on two different sampling rate of 22050Hz(input) and 24000Hz(output). Xtts [2] improved on tortoise tts by including more languages in the training of GPT2 with language embedding. VALLE [15] also introduced a language modeling approach to generate speech from text by using Neural codec [4] to discretize speech. Language modeling of text to speech gained a lot of attention and delivered the results but mainly in English and or Chinese. Given that the indic languages are not being represented in an unifying TTS system, limiting the application of tts to low resource indic languages. In this paper we aim to solve for this problem by collecting a massive indic dataset in 22 languages including (Assamese (in), Bengali (in), Bhojpuri (in), Bodo (in), Dogri (in), Odia (in), English (en), French (fr), Gujarati (in), German (de), Hindi (in), Italian (it), Kannada (in), Malayalam (in), Marathi (in), Telugu (in), Punjabi (in), Rajasthani (in), Sanskrit (in), Spanish (es), Tamil (in), Telugu (in)) as in Table 1 and training a TTS inspired by tortoise tts. The main contributions of our paper are:

- Training on 20k hours of Indic datasets across **22** languages.

- The model can perform **cross-lingual** text-to-speech synthesis out of the box.

- Architectural improvements using **semantic tokens**, a **Gemma-based backbone**, and a **flow matching model** instead of diffusion.

- **Model and code are publicly available** for reproducibility and further research.

## 2. Proposed Approach

MahaTTS can be divided into two components, namely M1 (text to semantics) and M2 (semantics to acoustic). Both Models are trained independently, and do not require joint training.

An illustration of our model is presented in Figure 1. Below, we present our technical approach in more detail.

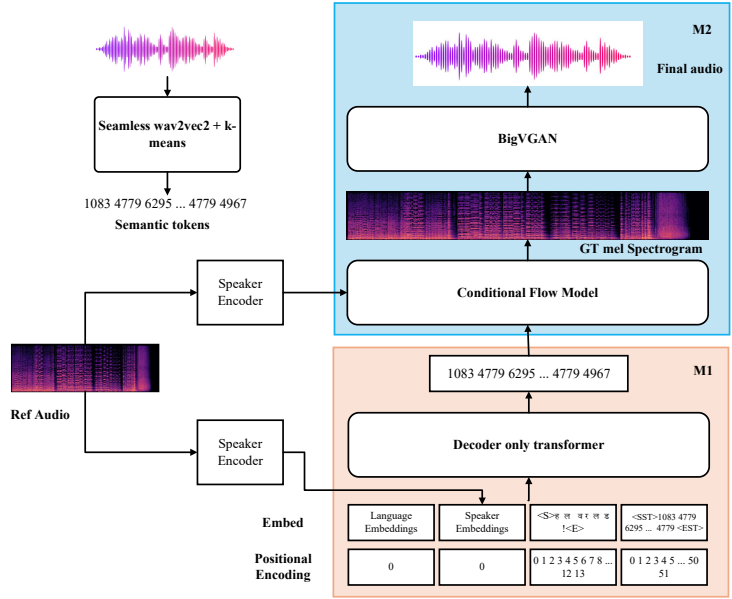

Figure 1. MahaTTSv2 Architechture

### 2.1. Speech Representations

Speech can be divided into two components, **Semantics** which cotains the linguistics of the speech like the content, intonation, emotion, and **Acoustics** which include speaker identity and other acoustic components. Tortoise TTS used VQVAE to extract quantized embeddings that have 8192 vectors. These vectors represent the abstract phonemes. In VQVAE as in (Tortoise TTS) and Wav2vec2 embeddings + k-means [7] are two popular choices for semantic representation of audio. VQVAE provides a quantized vector dictionary of 8192 vectors/tokens when trained on speech data; these vectors represent the abstract phonemes, 25 semantic tokens per second, however this requires joint training with M2 for a better performance. A pretrained Wav2Vec2 model like xlsr-1b [12] (trained on 436k hours of unlabelled speech across 128 languages), extracted embedding and fitting a k-means on top of it results in a 10000 dictionary of discrete phoneme sound. We have decided to go with this model as it does not require training and works out of the box for unseen languages when decoded [6], it does not require joint training.

### 2.2. M1

M1 maps tokenized text to Semantic tokens, this can be achieved through an autoregressive encoder-decoder/decoder-only model, or a non autoregressive model with a duration predictor. We choose a 0.5B gemma [3] based decoder-only model and train it directly on tokenized text and semantic tokens. Trained in a language modeling fashion with 0.1 loss weight to text tokens and 1.0 loss weight for semantics tokens, using different classification heads for text and semantics tokens. Transformer based decoder only models are very sensitive to positional encoding (ROPE [13] in our case), if the model has seen very less or no samples of small sentences or single words during training, it will not be able to produce single words during inference.

### 2.3. M2

M2 has two models, a Conditional Flow matching model [9] to map semantic tokens to MelSpectogram and BigVGAN to map the MelSpectogram to generate an audio waveform. Flow Model is a 300M flat alternating attention and convolutions layers with residual connections. The model is conditioned on semantic tokens and nearest 3 reference clips which are fed to a speaker encoder which trains along with the same model. The model is trained on 10 seconds speech audio samples and generalized well up to 30s speech generation due to RelativePosition-Bias as positional encoding. The audio speech is sampled at 24KHz. The model $\mathbf{v_t}$ predicts a vector field $\mathbf{u_t}$, which helps traverse a probability path $\mathbf{p_t}$ from a standard normal distribution $\mathcal{N}(\mathbf{0}, \mathbf{1})$ to the real data distribution. The expected FM loss is defined as:

$$\mathbf{L}_{\text{FM}} = \mathbf{E}_{t,\, p_t(x)} \left\| \mathbf{v_t}(\mathbf{x}) - \mathbf{u_t}(\mathbf{x}) \right\|^2 \quad (1)$$

Since we do not have direct access to $\mathbf{p_t}$ and $\mathbf{u_t}$,

we approximate them with a conditional probability path:

$$p_t = \mathbf{N}\left(x \mid \mu_t(\mathbf{x_1}), \sigma_t(\mathbf{x_1})^2 \mathbf{I}\right)$$

This approximation is valid because it yields gradients identical to (1) with respect to $\theta$

$$\mathbf{L}_{\text{CFM}} = \mathbf{E}_{t,\,\mathbf{p_t(x)}} \left\| \mathbf{v_t}(\mathbf{x} \mid \mathbf{x_1}) - \mathbf{u_t}(\mathbf{x} \mid \mathbf{x_1}) \right\|^2 \quad (2)$$

Using optimal transport (OT) as the flow:

$$\mathbf{p_t(x)} = (1-t)\mathbf{x_0} + t\mathbf{x_1}, \quad \mathbf{x_1}\,training data$$

At $t = 0$, $\mathbf{p_0}(\mathbf{x}) = x_0 =$ Noise, and at $t = 1$, $\mathbf{p_1}(\mathbf{x}) = \mathbf{x_1} =$ Data. The vector field becomes:

$$\mathbf{u_t(x)} = \frac{d}{dt}\mathbf{p_t(x)} = \mathbf{x_1} - \mathbf{x_0}$$

$$\mathbf{L}_{\text{OTCFM}} = \mathbb{E}\left\| \mathbf{v_t}\left((1-t)\mathbf{x_0} + t\mathbf{x_1} \mid y, \text{ref}\right) - (\mathbf{x_1} - \mathbf{x_0})\right\|^2 \quad (3)$$

Where:

- $x_1 \rightarrow$ Mini-batch of mel-spectrograms
- $x_0 \rightarrow$ Mini-batch of noise
- $y \rightarrow$ Semantic tokens
- ref $\rightarrow$ Reference file for speaker identity

A pre-trained BigVGAN [8] is used to convert the generated mel spectrograms to audio waveforms during inference. The M2 system in itself allows to act as a voice conversion model, an input speech can be converted into semantic tokens and can be generated with other speaker conditioning.

## 3. Dataset

We collected the open source datasets, accounting for 20K hours in 22 languages. We also recorded a proprietary dataset in 15 languages for over 100 hours used in training. Internally we have collected over 150k hours across 33 languages to train our future models. We tried to keep the length of the data samples in line with the normal distribution, as it is needed for the gemma model to be seen during training for positional encoding.

Data acquisition is a tedious process that involves a step-by-step process to procure and clean the data before using them for training purposes. We used VAD model to detect the audio with max continuous speech of 30s and min of 1s, to balance the distribution of the audio lengths, we made a recursive

| Languages | Total Hours | Percentage |
|---|---:|---:|
| assamese | 48.50 | 0.24% |
| bengali | 419.23 | 2.03% |
| bodo | 26.63 | 0.13% |
| dogri | 8.45 | 0.04% |
| english | 12038.88 | 58.36% |
| french | 1007.44 | 4.88% |
| german | 1514.56 | 7.34% |
| gujarati | 292.22 | 1.42% |
| hindi | 2265.75 | 10.98% |
| kannada | 40.66 | 0.20% |
| malayalam | 23.60 | 0.11% |
| manipuri | 27.10 | 0.13% |
| marathi | 873.28 | 4.23% |
| odia | 430.33 | 2.09% |
| punjabi | 80.03 | 0.39% |
| rajasthani | 20.05 | 0.10% |
| sanskrit | 84.01 | 0.41% |
| spanish | 889.68 | 4.31% |
| tamil | 29.37 | 0.14% |
| telugu | 357.79 | 1.73% |
| urdu | 152.35 | 0.74% |
| **Grand Total** | **20629.91** | **100.00%** |

Table 1. Dataset Composition across languages

VAD pipeline which gives normal duration samples. These samples are then passed to a speech quality evaluation model which will classify if the speech has any background noise, multiple speakers, music, clean speech, etc. We will take only clean processed speech and transcribe it into two different ASRs, and if the distance between the two outputs is less than a certain threshold, we would consider it for training. We would run a script to attend to the textual representation and clean the text, for example, in the case of Hindi most ASR are not trained to give out nuktas ("क़", "ख़", "ग़", "ज़", etc), we created a dictionary to map the nonnuktas words to nuktas. Many TTS systems still give probabilistic results for nuktas.

## 4. Experimental Setup

### 4.1. Pretraining

We augmented the data to 50k hours to increase the context window (30s) for the Gemma model. The model is trained on 8 A100 40gb, with a batch size of 1024 in bf16 for 7 epochs. We used different dataloaders based on the length of the sequences to efficiently use the GPUs for training. We used AdamW with 5e-5 LR and 1e-3 Weight decay. Adding a new language is fairly easy, the model picks up a new language really well with 200hours. The M1 model conditions on the similar nearest 3 reference clips to the target clip to get the style/intonation information, the clips are converted to melspectogram and passed through speaker encoder (six attention layers) and are then averaged over the time axis to produce 1024 fixed embedding, which are then averaged for the nearest three ref files; this is called speaker embedding. The model takes Language Embedding, Speaker Embedding, and Text tokens as input during inference and generates speech tokens to feed into the M2 system. The total vocabulary size is 10001 (semantic tokens)+ 4893 (character tokens) = 14894. The audio speech is sampled at 24KHz.

| Language | mahatts v2 (ours) | indic-parler-tts | indicF5 | veena | xtts |
|---|---|---|---|---|---|
| assamese | 42% | 80% | 54% | - | - |
| bengali | 7% | 39% | 3% | - | - |
| english | 2% | 2% | - | 3% | 0% |
| french | 18% | - | - | - | 17% |
| german | 15% | - | - | - | 75% |
| gujarati | 36% | 30% | 23% | - | - |
| hindi | 4% | 6% | 10% | 7% | 42% |
| italian | 2% | - | - | - | 25% |
| kannada | 16% | 57% | 14% | - | - |
| malayalam | 11% | 63% | 16% | - | - |
| marathi | 6% | 22% | 13% | - | - |
| odia | 73% | 40% | 43% | - | - |
| punjabi | 37% | - | 7% | - | - |
| sanskrit | 26% | 51% | - | - | - |
| spanish | 6% | - | - | - | 10% |
| tamil | 29% | 18% | 10% | - | - |
| telugu | 37% | 46% | 44% | - | - |

Table 2. WER (%) comparison across TTS systems for multiple languages. Lower is better; best scores are shaded green.

In the case of the Flow model, it was trained on 4 A5000s on 1k hours of data subset across all the languages with a combined batch size of 1024 samples, over 700k iterations. We used AdamW with 1e-4 LR and 1e-3 Weight decay.

### 4.2. Finetuning

While Finetuning on a new speaker Gemma model results in hallucination. We tried freezing the speaker encoder; it results in less hallucination but still more than the pre-trained model. Finally, freezing the classification heads for both text and semantic tokens resulted in comparable performance with the pre-trained model in terms of stability. Fine-tuned models result in better quality overall. For the flow model, it takes roughly 10k iterations, on a smaller batch size of 64 samples to fine-tune for a good quality output.

## 5. Evaluation
### 5.1. WER Evaluation

We also looked at generating 10 sentences for each language by different TTS systems and used scribe api to transcribe them back and calculated for the word error rate. The results for the similar are show in Table 4. Most of the errors are related to hallucinations with pronunciations issues following. We have shared the tested sentences in the code.

## 6. Conclusion  Future Work

The Shogun models work well for the Indic languages, and do pretty fine when finetuned. It supports 14 languages and works cross lingual out of the box. The code and model are made public.

We have internally collected 150k hours of data across 34 languages (25 indic), and are in process of collecting more data. The M1 models require more different conditionings like expressiveness and pace control. The M2 system requires a scaled training of the flow model as the infilling task for it to be sure zero shot as done in seamless [12].

### Acknowledgments

We thank the transformers library, MatchaTTs [10] codebase for inspiring the training code.